\documentclass[12pt]{article}
\pdfoutput=1
\usepackage{amsmath, amsfonts,euscript,bbm,mathrsfs,amssymb}
\usepackage{epsfig, cite}
\usepackage{color}
\usepackage{ifthen}
\usepackage{graphicx}
\usepackage{setspace} 
\usepackage{enumerate}

\newcommand{\ud}{\mathrm{d}}

\newcommand{\skyp}[1]{}

\begin{document}

\bigskip
\bigskip\bigskip\bigskip\bigskip\bigskip
\bigskip\bigskip\bigskip\bigskip\bigskip

\centerline{\Huge Invisibility cloaking without}
\centerline{\Huge superluminal propagation}
\bigskip
\bigskip
\bigskip
\centerline{\bf Janos Perczel$^{1a}$, Tomas Tyc$^{2b}$, and Ulf Leonhardt$^{1c}$}
\medskip
\centerline{$^1$School of Physics and Astronomy,}
\centerline{University of St Andrews,}
\centerline{North Haugh, St Andrews, KY16 9SS, UK}
\medskip
\centerline{$^2$Faculty of Science, Kotlarska 2, and} 
\centerline{Faculty of Informatics, Botanicka 68a,}
\centerline{Masaryk University, 61137 Brno, Czech Republic}
\medskip
\centerline{\it $^a$jp394@st-andrews.ac.uk}
\centerline{\it $^b$tomtyc@physics.muni.cz}
\centerline{\it $^c$ulf@st-andrews.ac.uk}

\bigskip
\bigskip
\bigskip
\bigskip
\bigskip
\bigskip
\begin{abstract}
 
Conventional cloaking based on Euclidean transformation optics requires that the speed of light should tend to infinity on the inner surface of the cloak. Non-Euclidean cloaking still needed media with superluminal propagation. Here we show by giving an example that this is no longer necessary.

\end{abstract}
\newpage


\newpage

\section{Introduction}

The development of transformation optics \cite{Service:2010,Dolin:1961,Greenleaf:2003,Leonhardt:2006a,Pendry:2006,Leonhardt:2006b,Shalaev:2008,Leonhardt:2009c,Leonhardt:2010,Chen:2010} and metamaterials \cite{Milton:2002,Smith:2004,Costas:2007,Sarychev:2007,Cai:2009,Capolino:2009} had lead to the theoretical prediction of invisible dipoles \cite{Alu:2005,Milton:2006} and invisibility cloaks \cite{Leonhardt:2006a,Pendry:2006}. Since then the principles behind invisibility have been demonstrated in a number of experiments that feature cloaking in some reduced form. Electromagnetic cloaking was first realised for microwaves of one frequency and polarisation \cite{Schuring:2006}. Numerous subsequent electromagnetic experiments achieved ``carpet cloaking'' \cite {Li:2008,Liu:2009,Valentine:2009,Gabrielli:2009,Ergin:2010,Chen:2011a,Zhang:2011a}. A carpet cloak does not make things invisible, but makes them appear to be flat. Approximate cloaking has also been realised through tapered waveguides \cite{Smolyaninov:2009,Tretyakov:2009}. Invisibility by plasmonic covering \cite{Alu:2005} was demonstrated as well \cite{Edwards:2009}. Furthermore, non-electromagnetic forms of cloaking have been demonstrated for acoustic waves \cite{Chen:2007,Zhang:2011b}.\\
However, the realisation of electromagnetic cloaking suffers from a practical and a fundamental problem \cite{Leonhardt:2009,Leonhardt:2009b,Chen:2011b}. The practical problem is that the material requirements for electromagnetic cloaking are very difficult to meet, whereas the fundamental problem is that perfect invisibility \cite{Pendry:2006} requires that light should propagate in certain cloaking regions with a superluminal phase velocity that tends to infinity. In principle, this can be achieved by metamaterials, but only for discrete frequencies that correspond to the resonant frequencies of the cloaking material \cite{Leonhardt:2006b,Chen:2011b}. In contrast, acoustic cloaking \cite{Chen:2007,Zhang:2011b} is much easier to achieve, since acoustic waves are not effected by relativistic causality and thus they are not restricted by their velocity in air.\\
Broadband cloaking based on non-Euclidean geometries has been proposed in \cite{Leonhardt:2009,Leonhardt:2009b} to avoid the requirement for infinite light velocities for electromagnetic cloaking. In this proposal the speed of light is finite in the entire cloaking region and, therefore, this cloak has the potential of working for a broad range of frequencies. However a  fundamental problem still remains. Since space has to be expanded to make room for the invisible region within the cloak, the implementation of this device will still demand superluminal propagation (i.e. propagation with a velocity that exceeds the speed of light in vacuum). The same is true for ``carpet cloaking'' \cite {Li:2008,Liu:2009,Valentine:2009,Gabrielli:2009,Ergin:2010,Chen:2011a,Zhang:2011a} where the velocity of light in the cloaking device must exceed the speed of light in the host material, i.e. vacuum if such devices were to find practical applications.  \\
In this paper we give an example of a device that achieves complete electromagnetic cloaking---not just ``carpet cloaking''---while all light velocities within the cloak are finite and less than the speed of light\footnote{The preprint \cite{Xie} proposes a different method for cloaking without superluminal propagation.}. Through this example we demonstrate that invisibility cloaking is possible without superluminal propagation and anomalous material requirements.

\section{Problem}
\begin{figure}[h!]
\begin{center}
\includegraphics[width=7cm]{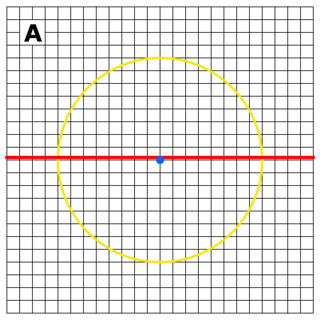}
\includegraphics[width=7cm]{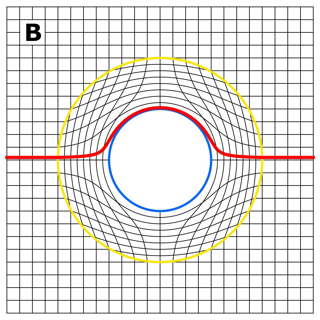}
\caption{A Euclidean cloaking device expands a single point in virtual space (blue dot in A) into an extended region in physical space (blue circle in B) through the curved transformation of the virtual coordinate grid. A light ray that is smoothly guided around the invisible region in physical space (B) appears to have passed through empty space (A), making the region within the blue circle (B) invisible. However, along the blue circle the phase velocity of light goes to infinity.}
\end{center}
\end{figure}
Since transformation optics establishes a one-to-one correspondence between spatial geometries and dielectric media, invisibility can be visualised in terms of virtual geometries. Therefore, we explain the main ideas of this paper in terms of pictures. The complete calculations behind these pictures can be found in the Appendix.\\
First, let us contrast our method to conventional cloaking based on coordinate transformation. Figure~1 explains the conventional cloaking method described in \cite{Pendry:2006}. A single point in virtual space -- also called electromagnetic or optical space -- is expanded into a finite region in physical space. The expansion is effected by the curved transformation of the straight virtual coordinate lines. Consequently, a light ray that traces out a curved trajectory in physical space (Fig.~1B) appears to propagate along a straight line in virtual space (red line in Fig.~1A). This creates the illusion that light has propagated through empty space, making the central region in physical space (region bounded by the blue circle in Fig.~1B) invisible.\\
However, there is a fundamental problem with this device. Light crosses the central point in virtual space (blue dot in Fig.~1A) in an infinitely short time. This point turns into an extended region in physical space, which still has to be traversed in an infinitely short instant. Therefore, the phase velocity of light will tend to infinity along the outer boundary of the invisible region in physical space. Infinite light velocities can be achieved by metamaterials, but only for discrete resonant frequencies of the material \cite{Chen:2011b,Leonhardt:2009}, fundamentally restricting the applicability of this cloak.\\
In contrast, the broadband non-Euclidean invisibility device proposed in \cite{Leonhardt:2009} achieves full cloaking without the need for infinite light speeds. It makes use of the rich structure of non-Euclidean geometries to avoid infinite expansions and singularities.\\
The background geometry of this cloak is established through the expansion of space, which is similar to optical conformal mapping \cite{Leonhardt:2006a,Leonhardt:2006c,Leonhardt:2010,Chen:2011b} (see Appendix \ref{Ap:Branches}). A finite region in physical space (red eye-shaped region in Fig.~2A) is turned ``inside out'' to produce a hidden infinite virtual plane (red lower plane in Fig.~2B). The blue outer region in physical space (Fig.~2A)  is stretched to make the upper sheet complete without the red region (Fig.~2B).
\begin{figure}[h!]
\begin{center}
\includegraphics[width=9.7cm]{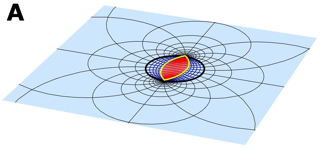}\\[0.6cm]
\includegraphics[width=9.7cm]{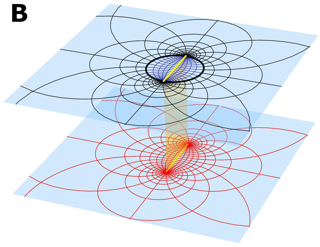}
\caption{The expansion of space. The red eye-shaped region in physical space (A) is expanded into a red infinite virtual plane (B). The blue region (A) is stretched to make the upper sheet complete without the red region (B). A light ray propagating with the speed of light along the straight yellow virtual branchcut (B) will have to follow the longer curved yellow line in physical space (A). Therefore, in physical space light will have to propagate faster than the speed of light in vacuum.}
\end{center}
\end{figure}
The two sheets remain connected by the yellow branchcut (Fig.~2B), through which light can pass from one sheet to another. The two virtual sheets provide the background geometry of the device. By the appropriate modification of the lower sheet, objects placed there can be hidden from sight. For the time being we will focus our attention on the blue region of the upper sheet, which we shall call the ``stretched region'' of the cloak.\\
It is clear from Fig.~2 that the blue region was stretched everywhere by merely finite factors. Therefore, establishing this geometry will not require infinite light speeds in contrast to the previous cloaking example (Fig.~1). However, a problem still remains. A light ray that propagates along the straight yellow branchcut in virtual space (Fig.~2B) will have to trace out a slightly longer curved path along the physical branchcut (Fig.~2A) in the same amount of time. Since light travels in virtual space with the speed of light in vacuum \cite{Leonhardt:2009}, the phase velocity in physical space will have to be faster than $c$. This can be achieved only for certain frequency bands that are around the resonant frequencies of the material \cite{Milonni:2005}, which restricts the applicability of the cloak. 
\section{Solution}
\addtocontents{toc}{\protect\setcounter{tocdepth}{1}}

\subsection{Geometry}

In the previous section we saw that the Non-Euclidean Cloak requires that light should propagate faster than the speed of light in vacuum. In order to avoid such anomalous light velocities, we construct a device that makes use of the same space expansion as the Non-Euclidean Cloak (Fig.~2), but is built up from different constituents.\\
To slow the light rays down in the stretched region of the cloak (blue region in Fig.~3A), we place an optically dense medium (that is, a medium with high refractive index) on the upper sheet of virtual space (Fig.~3B). This medium must itself be invisible to avoid revealing the presence of the invisibility device. Therefore, we use the partially transmuted Invisible Sphere as the background profile for the upper sheet of virtual space (purple circle in Fig.~3B). All rays entering this Sphere perform a complete loop in it and emerge with their original direction restored (Fig.~3B). Therefore, the entire sphere appears invisible. At the same time, if a light ray crosses the yellow branchcut while propagating through the Sphere, it will pass onto the lower virtual sheet of the geometry (Fig.~3B). We place Maxwell's Fish Eye Lens on the lower virtual sheet and surround it by a circular mirror (dark cyan circle in Fig.~3B) to guide all light rays that enter the lower sheet back to the yellow branchcut. The light ray reemerges from the branchcut on the upper sheet and finishes its trajectory in the Invisible Sphere (Fig.~3B). Since the original direction of the ray is restored when it emerges from the device, it appears to have propagated through empty space. Note that there is a region outside the mirror on the lower sheet that is inaccessible to light. Any object placed in that region is undetectable to light and thus becomes invisible.
\begin{figure}[h!]
\begin{center}
\includegraphics[width=9.5cm]{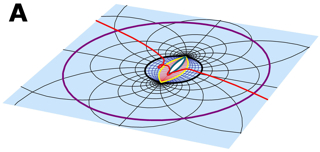}\\[0.6cm]
\includegraphics[width=9.5cm]{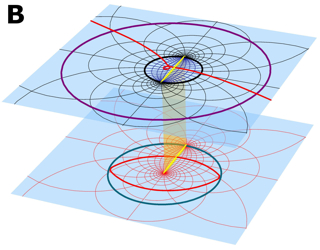}
\caption{Light propagation in physical space (A) and virtual space (B). We place the partially transmuted Invisible Sphere on the upper sheet of our virtual geometry (region bounded by the purple circle in B) and put Maxwell's Fish Eye Lens on the lower virtual sheet and surround it by a circular mirror (dark cyan circle in B). Light explores only the interior of the mirror in virtual space (B). The unexplored infinite region of the lower virtual sheet (B) transforms into a finite region in physical space (white spindle-shaped region in A) that is unaccessible to light and where objects can be hidden. In physical space light completes a loop in the device, bounces off the mirror twice and leaves the device with its original direction restored.}
\end{center}
\end{figure}
Fig.~3A shows the light trajectory in physical space. The light ray (red) completes a loop that is characteristic of the Invisible Sphere, bounces off the spindle-shaped mirror (dark cyan) twice and leaves the device with its original direction restored. Note that the infinite region outside the circular mirror on the lower virtual plane (Fig.~3B) transforms into a finite spindle-shaped region in physical space (white region in Fig.~3A). This region is inaccessible to light and any object placed in there remains hidden from sight. Since the stretched region of the cloak now sits in the optically dense Invisible Sphere, all light rays within it are propagating slower than the speed of light.\\

\subsection{Building blocks} \label{Sec:BuildingBlocks}
\addtocontents{toc}{\protect\setcounter{tocdepth}{2}}

In this section we explain the partially transmuted Invisible Sphere and Maxwell's Fish Eye Lens with the mirror, which serve as the two fundamental building blocks of our virtual geometry.\\
The Invisible Sphere is an optical lens that is invisible in the limit of geometrical optics \cite{Hendi:2006,Leonhardt:2010}. Each incident ray completes a loop in the Sphere and emerges in the direction of its entry (Fig.~4A), creating the illusion that space is empty.\\

\begin{figure}[h!]
\begin{center}
\includegraphics[width=7cm]{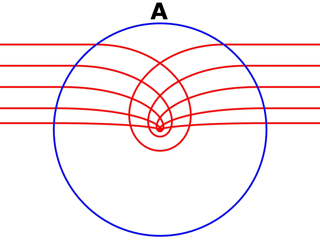}
\includegraphics[width=7cm]{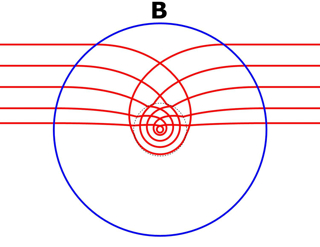}
\caption{The Invisible Sphere and its partial transmutation. Each ray entering the Invisible Sphere performs a loop in it and leaves in the direction of its entry (A), making the Sphere invisible. The singularity at the centre of the Invisible Sphere can be removed by expanding space around the centre within a given radius (B).}
\end{center}
\end{figure}
This device has a spherically symmetric refractive index profile given by \cite{Minano:2006}

\begin{eqnarray}\label{eq:ISprofile1}
n(r)=\bigg( Q-\,1/(3Q)\bigg)^2 \:\:  \text{if $r \leqslant 1$, }   \text{and} \:\:
n(r)=1 \:\: \text{ if $r > 1$,}
\end{eqnarray}

where

\begin{equation}\label{eq:ISprofile2}
Q(r)=\sqrt[3]{-\frac{1}{r}+\sqrt[2]{\frac{1}{r^2}+\frac{1}{27}}}\text{.}
\end{equation}

In the expression above we assume that the device has unit radius. It is easy to see that $n(r)$ tends to infinity as $r \to 0$. This means that the speed of light goes to zero as it gets close to the geometric centre of the device. This is extremely difficult to achieve in practice.\\
Therefore, we make use of partial transmutation \cite{PartialTransmutation:2011}---a modification of the method presented in \cite{Tyc:2008}---to expand space with a non-constant factor in the immediate vicinity of the singularity within a given radius of the device (dashed circle in Fig.~4B). By this expansion we give more space for the propagation of light and thus ensure that the refractive index is finite everywhere (i.e. light propagates with a non-zero velocity) within the entire device \cite{Tyc:2008,PartialTransmutation:2011} (see Appendix \ref{Ap:InvisibleSphere}).\\
Maxwell's Fish Eye Lens \cite{Maxwell:1854,Luneburg:1964} is an optical device in which all light trajectories trace out perfect circles (Fig.~5A). It has a spherically symmetric refractive index profile given by

\begin{equation}\label{eq:Mprofile}
n(r)=\frac{2\,n_{l}}{1+(r/l)^{2}}\text{.}
\end{equation}

In the expression above $l$ is the radius of the ``equator'' of the device at which a circular mirror can be introduced (blue circle in Fig.~5B) such that all trajectories remain closed.\\

\begin{figure}[h!]
\begin{center}
\includegraphics[width=7cm]{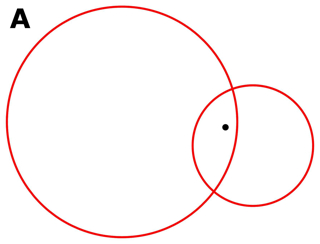}
\includegraphics[width=7cm]{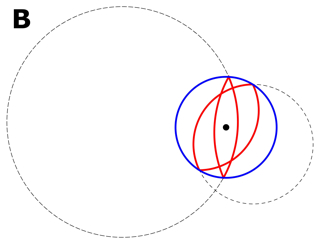}
\caption{Maxwell's Fish Eye Lens without and with the mirror. In Maxwell's Fish Eye profile all light trajectories are circular (A). When the Fish Eye Lens is surrounded by a circular mirror of appropriate radius, the trajectories still remain closed (B).}
\end{center}
\end{figure}
Here $n_{l}$ is the value of the refractive index at the equator.  It can assume any values that are greater than or equal to unity. A choice of large $n_{l}$ slows the light rays down, but it does not affect the trajectories overall, since the form of the trajectories in the device are dependent only on the relative distribution of the refractive indices. It is precisely this feature of the lens that we can exploit to avoid superluminal propagation in the inner branch of the cloak (see Appendix \ref{Ap:FishEye}).

\section{Materials}

Now we turn to calculating the material properties of the cloak in physical space. We make use of the method described in \cite{Leonhardt:2006b,Leonhardt:2009c,Leonhardt:2010}. We utilise the ideas of transformation optics \cite{Dolin:1961,Greenleaf:2003,Leonhardt:2006a,Pendry:2006,Leonhardt:2006b,Shalaev:2008,Leonhardt:2009c,Leonhardt:2010,Chen:2010}, which establish a connection between spatial geometries and dielectric media.
We calculate the permittivity and permeability tensors in bipolar coordinates (see Appendix \ref{Ap:MaterialProperties}). The tensors obtained have the following form:
\begin{equation}
\varepsilon^{i}_{\, k}=\mu^{i}_{\, k}=\text{diag}(\varepsilon_{\sigma},\varepsilon_{\tau},\varepsilon_{z})\,.
\end{equation}

These two tensors completely describe the electromagnetic response of a dielectric material and their eigenvalues (i.e. $\varepsilon_{\sigma}$, $\varepsilon_{\tau}$, and $\varepsilon_{z}$) can be engineered into metamaterials to implement the cloak. The tensor eigenvalues and the refractive index values are related as \cite{Leonhardt:2009c,Leonhardt:2010}

\begin{equation} \label{eq:refindex}
n_{\sigma}=\sqrt{\varepsilon_{\tau}\varepsilon_{z}}, \,\: n_{\tau}=\sqrt{\varepsilon_{\sigma}\varepsilon_{z}}, \,\: n_{z}=\sqrt{\varepsilon_{\sigma}\varepsilon_{\tau}} \, \text{.}
\end{equation}

It is clear from the expressions above that the refractive index values could be greater than one even if not all of the individual tensor eigenvalues are greater than unity. However, in order to avoid anomalous dispersion and absorption at all frequencies, it is important that the individual tensor eigenvalues are also greater than unity.\\
Through the appropriate positioning of the building blocks of our electromagnetic geometry and the adjustment of their parameters, all tensor eigenvalues can be brought above unity.
We calculated the distribution of the tensor eigenvalues $\varepsilon_{\sigma},\varepsilon_{\tau}$, and $\varepsilon_{z}$ for the geometry shown in~Fig.~3. The detailed calculations can be found in Appendix \ref{Ap:MaterialProperties}. We plotted the eigenvalue distributions in Fig.~6.\\

\begin{figure}[h!]
\begin{center}
\includegraphics[width=5.4cm]{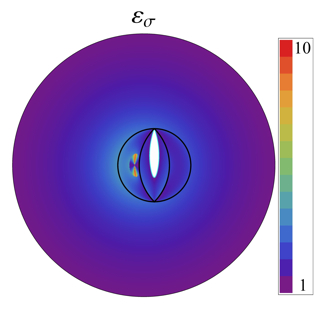}
\includegraphics[width=5.4cm]{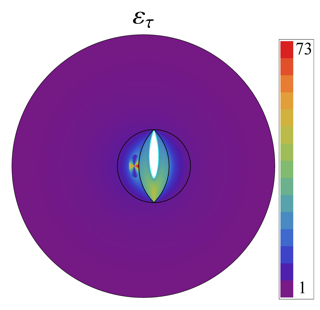}
\includegraphics[width=5.7cm]{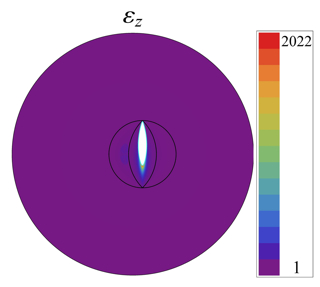}
\caption{Distribution of the tensor eigenvalues within the combined cloak. The boundaries of the stretched region of the cloak are shown in black within the partially transmuted Invisible Sphere. All tensor eigenvalues are greater than or equal to unity and finite.}
\end{center}
\end{figure}
We found that $\varepsilon_{\sigma} \in [1,9.428] $, $\varepsilon_{\tau} \in [1,73.21]$, and $\varepsilon_{z} \in [1,2012.12]$. It is clear that all values of the tensor components are now greater than or equal to one, and finite. Therefore, all refractive index components are also greater than or equal to unity [see equation \eqref{eq:refindex}].\\We note that some of these values are quite extreme and might be very difficult to achieve in practice. However, this cloak is just an example, which demonstrates that full invisibility cloaking is possible without superluminal velocities. The range of material parameters could be optimized by changing the topology of the cloak slightly (see Appendix \ref{Ap:Optimization}) or by introducing further coordinate transformations to eliminate regions with high permittivity and permeability values (see \cite{Tyc:2008} for an example).\\
Figure~6 shows that the profile of the cloak is highly anisotropic within the inner branch of the cloak (demarcated by black lines) and isotropic outside it. In Fig.~7 we depict a light trajectory against the values of the $\varepsilon_{\sigma}$ component in the presence of the mirror. The light ray performs a loop within the transmuted centre of the Invisible Sphere, bounces off the mirror twice and leaves the device with its original direction restored. Objects placed in the white region behind the mirror are invisible.\\
Finally, we note that even though this cloak is two-dimensional, the ideas behind it can be readily extended to construct three-dimensional version of it. A three-dimensional version of the non-singular expansion of space is given in \cite{Leonhardt:2009}, and Maxwell's Fish Eye as well as the partially transmuted Invisible Sphere are both three-dimensional profiles. Naturally, the parameter ranges will be different for the three-dimensional cloak, but through appropriate optimization superluminal propagation can be avoided in the entire cloak.

\begin{figure}[h!]
\begin{center}
\includegraphics[height=8.cm]{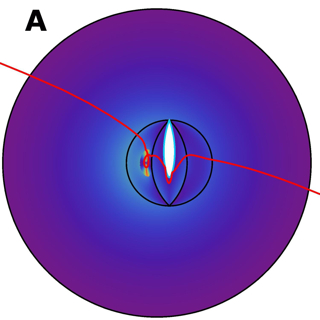}\\[1cm]
\includegraphics[height=8.cm]{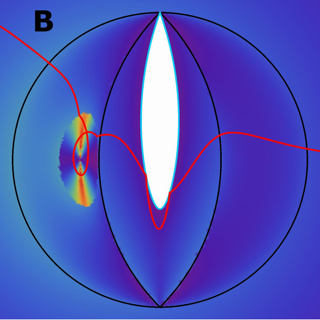}
\caption{A light trajectory is shown against the distribution of the $\varepsilon_{\sigma}$ values. The light ray enters the device, completes a loop, bounces off the mirror twice and leaves the cloak with its original direction restored (A). Figure B gives a closer view of the vicinity of the inner branch of the cloak. Objects placed within the white region are invisible.}
\end{center}
\end{figure}

\section{Conclusion}

In conclusion, we have combined the partially transmuted Invisible Sphere with Maxwell's Fish Eye Lens to obtain a cloak whose permittivity and permeability tensor eigenvalues are everywhere greater than or equal to unity and finite. This example demonstrates that full invisibility cloaking is possible without superluminal propagation and anomalous material requirements.

\section*{Acknowledgements}

We thank Natalia Korolkova and William Simpson for their valuable comments and help. Our work was supported by the Engineering and Physical Sciences Research Council and the Royal Society.

\appendix

\section{Appendix}

\subsection{Branches in bipolar coordinates} \label{Ap:Branches}

We use bipolar coordinates and the expansion of space to establish the two virtual sheets depicted in Fig.~2.
We parameterise physical space by bipolar coordinates, which can be related to the Cartesian coordinates by 

\begin{equation} \label{eq:bipolars}
x= \frac{a\,\text{sinh}\,\tau}{\text{cosh}\,\tau-\text{cos}\,\sigma }\,, \qquad y= \frac{a\,\text{sin}\,\sigma}{\text{cosh}\,\tau-\text{cos}\,\sigma }\, .
\end{equation}

In the expression above $a$ is the radius of the boundary of the stretched region of the cloak (i.e. half the length of the branchcut). Working in the complex plane we can rewrite these relations succinctly (equations (S29 and S30) in \cite{Leonhardt:2009b})

\begin{equation} \label{eq:complexbipolars}
x+\text{i}\,y=\text{i}\, a\,\text{cot}\bigg(\frac{\sigma+\text{i}\,\tau}{2}\bigg),
\end{equation}

with the inverse relation
\begin{equation} \label{eq:inversecomplexbipolars}
\sigma+\text{i}\,\tau=2\,\text{i}\,\text{arcoth}\bigg(\frac{x+\text{i}\,y}{a}\bigg).
\end{equation}

We map the physical plane (Fig.~2A) onto two virtual sheets (Fig.~2B) through the coordinate transformation given by (equation (S34) in \cite{Leonhardt:2009b}) 

\begin{equation} \label{eq:SigmaPrime}
\sigma'(\sigma)= \left\{
\begin{array}{lr}
\sigma & \text{for } |\sigma| \le \dfrac{\pi}{2}\qquad\:\:\\
\text{sgn}\,\sigma \bigg(\dfrac{4\sigma^{2}}{\pi}-3|\sigma|+\pi\bigg) \qquad\qquad\qquad& \text{for } \dfrac{\pi}{2}< |\sigma|\le\pi\,.
\end{array} \right.
\end{equation}

Here $\sigma'$ and $\tau$ are the new set of coordinates parameterising virtual space. While $\sigma$ runs from $-\pi$ to $\pi$, covering the plane once, $\sigma'(\sigma)$ runs from $-2\pi$ to $2\pi$ and covers the entire plane twice thus giving rise to two virtual planes (Fig.~2B). Outside the black circle in Fig.~2A~and~2B (corresponding to $|\sigma|\leq\pi/2$) $\sigma$ and $\sigma'$ agree since that region undergoes no transformation.\\
The two branches of virtual space are connected by the branchcut (yellow line in Fig.~2B), through which light can pass from one sheet onto another. In virtual space the bipolar and Cartesian coordinates are related by the expression

\begin{equation} \label{eq:inversecomplexbipolarsvirtual}
\sigma'+\text{i}\,\tau'=2\,\text{i}\,\text{arcoth}\bigg(\frac{x'+\text{i}\,y'}{a}\bigg).
\end{equation}

It is useful to note that  the expansion of space can be reversed using the inverse relation between $\sigma$ and $\sigma'$ (S35 in \cite{Leonhardt:2009b}):

\begin{equation} \label{eq:Sigma}
\sigma(\sigma')= \left\{
\begin{array}{lr}
\sigma' & \text{for } |\sigma'| \le \dfrac{\pi}{2}\qquad\quad\\
\dfrac{\text{sgn}\,\sigma'}{8} \bigg( 3\pi+\sqrt{16\pi|\sigma'|-7\pi^2} \bigg) \qquad\qquad\qquad& \text{for } \dfrac{\pi}{2}< |\sigma'|\le2\pi\:.
\end{array} \right.
\end{equation}

\subsection{Invisible Sphere}\label{Ap:InvisibleSphere}

The refractive index profile of the Invisible Sphere (Fig.~4A) is given by equations \eqref{eq:ISprofile1} and  \eqref{eq:ISprofile2} \cite{Minano:2006}. In these expressions we assume that the device has unit radius. It can be shown that $n(r)$ tends to infinity as $r\to 0$ and diverges as\, $\sim r^{-2/3}$ near the geometric centre of the sphere.\\
We remove the singularity at the centre by the method of partial transmutation \cite{PartialTransmutation:2011}. We expand space around the singularity within a finite radius $b$ to give enough space for the propagation of light. Thus we bring all refractive index values within a finite range. Our choice of $b$ for the transmutation will depend on the particular geometrical set-up used (see section \ref{Ap:Optimization} on optimization), but as \cite{PartialTransmutation:2011} shows, $b$ will always have to be smaller than 0.2489 to avoid tensor eigenvalues that are smaller than one.\\
Since $n(r)$ of the Invisible Sphere diverges as $\sim r^{-2/3}$, partial transmutation requires that we apply the following transformation to the radial coordinate:

\begin{equation} \label{eq:TransmutationFunction}
R(r)=\frac{1}{\mathcal{N}(b)} \, r^{p+1} \;\:  \text{if $r \leqslant b$, }  \: \text{and} \;\:
R(r)=r \;\: \text{ if $r > b$}\,.
\end{equation}
 
In the equation above $\mathcal{N}$ is the normalisation constant that ensures that the transformation is continuous at the boundary of the transmutation $r=b$. Therefore, it follows that $\mathcal{N}=b^{-2/3}$.\\
The material properties of the partially transmuted Invisible Sphere (Fig.~4B) -- that would mimic the coordinate transformation described above -- can be calculated in terms of spherical coordinates from the formula derived in \cite{Tyc:2008} (see also \cite{PartialTransmutation:2011}):

\begin{equation} \label{eq:TransmutationTensor}
\varepsilon^{i}_{\, k}=\mu^{i}_{\, k}=\text{diag}(\varepsilon_{R},\varepsilon_{\theta},\varepsilon_{\phi})=\text{diag}\Big( \frac{n \,r^2}{R^2}\frac{\ud R}{\ud r},\frac{n\,\ud r}{\ud R},\frac{n\,\ud r}{\ud R} \Big).
\end{equation}

In the above expression $n$ is the refractive index profile of the Invisible Sphere and $R$ is given by equation \eqref{eq:TransmutationFunction}.\\
For the particular partially transmuted Invisible Sphere positioned on the upper sheet of the virtual geometry in Fig.~3B, we chose the radius of the transmutation to be $b=0.075\,$ to keep all combined tensor eigenvalues of the cloak greater than one (see section \ref{Ap:Optimization} on optimization). We plotted the transmuted values of $\varepsilon_{R}$ and $\varepsilon_{\theta}=\varepsilon_{\phi}$ in Fig.~8. Clearly, all values are finite and greater than one. Note that the region within the radius of the transmutation is anisotropic, whereas the outer region has been unaffected by the transmutation and thus remains isotropic.\\
Also note that there is an alternative way of transmuting singularities without inducing anomalous material properties \cite{Danner:2010}. However, this method reduces the refractive index in the entire profile of the transmuted lens and, therefore, it is unsuitable for our purposes.
\begin{figure}[h!]
\begin{center}
\includegraphics[width=8.3cm]{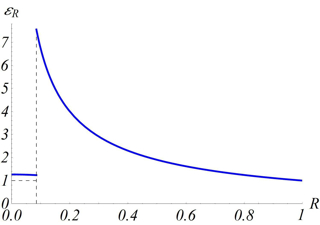}
\includegraphics[width=8.3cm]{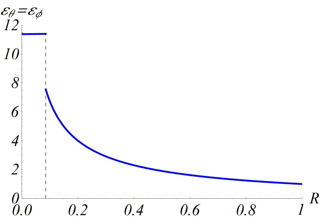}
\caption{Values of the radial and angular tensor components in the partially transmuted Invisible Sphere, where the radius of transmutation is $b=0.075$. All eigenvalues are greater than one and finite. The index profile of the Invisible Sphere outside the radius of the transmutation remains unaffected.}
\end{center}
\end{figure}

\subsection{Maxwell's Fish Eye and the mirror} \label{Ap:FishEye}

The refractive index profile of Maxwell's Fish Eye Lens \cite{Maxwell:1854,Luneburg:1964} (Fig.~5A) is given by
equation \eqref{eq:Mprofile}. We can solve Hamilton's equations numerically (see section \ref{Ap:HamiltonsEquations} on Hamilton's equations) to obtain the trajectories depicted in Fig.~5A.\\
Maxwell's Fish Eye can be surrounded with a circular mirror of radius $l$ (where $l$ is the radius of the `equator' of the device) such that all trajectories remain closed (Fig.~5B). To model the reflection of rays by the mirror, we express the light trajectories in the x-y plane in terms of the complex coordinate $z'(t)=x'(t)+\text{i}\,y'(t)$ and apply the transformation \cite{Leonhardt:2009b}

\begin{equation} \label{eq:MirrorFunction}
z'\to\dfrac{l}{(z'/l)^{*}}\quad \text{for}\quad|z'|>l.
\end{equation}

In the expression above $^*$ denotes complex conjugation.\\
For the particular Fish Eye Lens positioned on the lower sheet of the virtual geometry (Fig.~3B) $l=2a$, where $a$ is half the length of the branchcut (see sections \ref{Ap:Branches} and \ref{Ap:Optimization}), and $n_{l}$ was chosen to equal 5 to make sure that all tensor eigenvalues are greater than unity within the inner region of the cloak. The Fish Eye is centered around $(x',y')=(-a,0)$. This point corresponds to one end of the branchcut.

\subsection{Hamilton's equations} \label{Ap:HamiltonsEquations}

In this section we describe how to obtain the light trajectories in physical space. To get the trajectories in the outer branch of the cloak (blue and black region in Fig.~3A), we first solve Hamilton's equations in the Invisible Sphere (Fig.~4A). We mimic the effect of the partial transmutation by transforming the radial components of the trajectories according to equation \eqref{eq:TransmutationFunction} (Fig.~4B). Finally, we open up space within the Sphere along the branchcut using equation \eqref{eq:SigmaPrime}, which cuts up the trajectories into two separate parts and pushes them apart while the red region gets inserted in between them (Fig.~3A). The two separate parts of the trajectory in the blue region will be joined by the trajectory in the inner cloak.\\
To obtain the trajectories in the inner cloak (red region in Fig.~3A), we solve Hamilton's equations in Maxwell's Fisheye profile with $l=2a$ and $n_{l}=5$. The initial conditions for the solution of Hamilton's equations are provided by the trajectories in the outer cloak, where the incident wavevector has to be modified to take refraction into account at the branchcut (see section \ref{Ap:Optimization}). Next we shift the trajectories such that they are centred around one of the endpoints of the branchcut $(x',y')=(-a,0)$. We switch from Cartesian to bipolar coordinates using equation \eqref{eq:inversecomplexbipolarsvirtual}. Then we shift the values of $\sigma'$ from the intervals $[-\pi,0]$ and $[0,\pi]$ to $[\pi,2\pi]$ and $[-2\pi,-\pi]$ respectively and reverse the expansion of space using equation \eqref{eq:Sigma}. Thus we obtain the connected physical trajectories depicted in Fig.~3A and Fig.~7.\\
We find it convenient for ray tracing to rewrite Hamilton's equations in a complex form. Using the dispersion relation $\omega=ck/n$ and setting $c=1$, Hamilton's equations can be expressed as (equation (4.10) in \cite{Leonhardt:2010})
\begin{equation} \label{eq:HamiltonsEquations}
\dfrac{\text{d}\boldsymbol{r}}{\text{d}t}=\dfrac{1}{n}\dfrac{\boldsymbol{k}}{|\boldsymbol{k}|}\,,
\end{equation}
and
\begin{equation} \label{eq:HamiltonsEquations2}
\dfrac{\text{d}\boldsymbol{k}}{\text{d}t}=\dfrac{|\boldsymbol{k}|}{n^2}\,\nabla n\,,\quad \text{where}\quad   \nabla n=\dfrac{\partial n}{\partial|\boldsymbol{r}|}\dfrac{\boldsymbol{r}}{|\boldsymbol{r}|}\,.
\end{equation}
We rewrite equation \eqref{eq:HamiltonsEquations2} as
\begin{equation} \label{eq:HamiltonsEquations3}
\dfrac{\text{d}\boldsymbol{k}}{\text{d}t}=|\boldsymbol{k}|\,\text{g}(|\boldsymbol{r}|)\,\dfrac{\boldsymbol{r}}{|\boldsymbol{r}|}\quad\: \text{with} \quad\: \text{g}(|\boldsymbol{r}|)=\dfrac{1}{n^2}\dfrac{\partial n}{\partial|\boldsymbol{r}|}\,.
\end{equation}
We solve equations \eqref{eq:HamiltonsEquations} and \eqref{eq:HamiltonsEquations3} in terms of the complex position and momentum coordinates
\begin{equation} \label{eq:ComplexCoordinatesHE}
z(t)=x(t)+\text{i}\,y(t) \qquad \text{and} \qquad k(t)=k_{x}(t)+\text{i}\,k_{y}(t)\,.
\end{equation}
Therefore, we substitute $z(t)$ for $\boldsymbol{r}$ and $k(t)$ for $\boldsymbol{k}$ in equantions \eqref{eq:HamiltonsEquations} and \eqref{eq:HamiltonsEquations3} and thus we obtain the complex form of Hamilton's equations
\begin{equation} \label{eq:ComplexHamiltonsEquations}
\dfrac{\text{d}z}{\text{d}t}=\dfrac{1}{n(|z|)}\dfrac{k}{|k|}\,, \qquad \dfrac{\text{d}k}{\text{d}t}=|k|\,\text{g}(|z|)\,\dfrac{z}{|z|},\quad \text{with} \quad \text{g}(|z|)=\dfrac{1}{n^2}\dfrac{\partial n}{\partial|z|}\,.
\end{equation}

These equations are numerically stable.

\subsection{Calculation of material properties} \label{Ap:MaterialProperties}

The material properties of the cloak can be calculated using the recipe described in \cite{Leonhardt:2006b,Leonhardt:2009c,Leonhardt:2010}. We proceed in the following steps:
(i) we write down the line element $\ud s$ that describes our virtual geometry, (ii) express $\ud s$ in terms of the coordinates that parameterise physical space ($\sigma$, $\tau$, and $z$), and (iii) read off the metric tensor $g_{ij}$ of the geometry from $\ud s$. Finally, (iv) we calculate the permittivity and permeability eigenvalues from $g_{ij}$ using equations (8) and (9) from \cite{Leonhardt:2006b}:
\begin{equation} \label{eq:Recipe}
\varepsilon^{i}_{\:k}=\mu^{i}_{\:k}=\frac{\sqrt{g}}{\sqrt{\gamma}}g^{ij}\gamma_{jk}\,.
\end{equation} 
In the equation above $g^{ij}$ is the inverse of $g_{ij}$, $g$ is the determinant of $g_{ij}$, $\gamma_{jk}$ is the metric of the background geometry (here: empty space described by bipolar coordinates) and $\gamma$ is the determinant of $\gamma_{ij}$. In all our calculations we assume that the device is impedance matched to vacuum and so $\varepsilon^{i}_{\:k}=\mu^{i}_{\:k}$ \cite{Leonhardt:2006b}.\\
The cloak has two different branches that undergo different transformations, therefore, we will treat them separately.

\subsubsection{Outer branch} \label{Ap:OuterBranch}

First, we calculate the material properties of the outer branch of the cloak (black and blue region in Fig.~3A), where $\pi/2<|\sigma|\le3\pi/4$. This physical region corresponds to the upper sheet of the virtual geometry (Fig.~3B). The upper sheet of the device carries the refractive index profile of the Invisible Sphere $n_{\scriptscriptstyle \text{IS}}$. Therefore, the line element in spherical coordinates is given by:
\begin{equation}
\ud s^2=n_{\scriptscriptstyle \text{IS}}^2\big(\ud r^2+r^2\,\ud\theta^2+r^2\text{sin}^2\theta\,\ud\phi^2\big)\,.
\end{equation}
We perform the radial expansion given by equation \eqref{eq:TransmutationFunction}, which corresponds to partial transmutation
\begin{equation}
\ud s^2=n_{\scriptscriptstyle \text{IS}}^2\bigg[\bigg(\dfrac{\ud r}{\ud R}\bigg)^2\ud R^2+r^2\,\ud\theta^2+r^2\text{sin}^2\theta\,\ud\phi^2\bigg]\,.
\end{equation}
Next we change from spherical to Cartesian coordinates
\begin{eqnarray}
\ud s^2=n_{\scriptscriptstyle \text{IS}}^2 \Bigg\{
    \Bigg[\bigg(\dfrac{\ud r}{\ud R}\bigg)^2\bigg(\dfrac{\ud R}{\ud x}\bigg)^2+r^2\,\bigg(\dfrac{\ud \theta}{\ud x}\bigg)^2+r^2\text{sin}^2\theta\,\bigg(\dfrac{\ud \phi}{\ud x}\bigg)^2\Bigg]\,\ud x^2 \;\quad \nonumber \\
+ \Bigg[\bigg(\dfrac{\ud r}{\ud R}\bigg)^2\bigg(\dfrac{\ud R}{\ud y}\bigg)^2+r^2\,\bigg(\dfrac{\ud \theta}{\ud y}\bigg)^2+r^2\text{sin}^2\theta\,\bigg(\dfrac{\ud \phi}{\ud y}\bigg)^2\Bigg]\,\ud y^2 \;\quad\nonumber \\
+ \Bigg[\bigg(\dfrac{\ud r}{\ud R}\bigg)^2\bigg(\dfrac{\ud R}{\ud z}\bigg)^2+r^2\,\bigg(\dfrac{\ud \theta}{\ud z}\bigg)^2+r^2\text{sin}^2\theta\,\bigg(\dfrac{\ud \phi}{\ud z}\bigg)^2\Bigg]\,\ud z^2\Bigg \} .
\end{eqnarray}
We define the expressions in square brackets in front of $\ud x^2$, $\ud y^2$, and $\ud z^2$ as $\epsilon_{x}$, $\epsilon_{y}$, and $\epsilon_{z}$ respectively and obtain
\begin{equation}
\ud s^2=n_{\scriptscriptstyle \text{IS}}^2\big(\epsilon_{x}\,\ud x^2+\epsilon_{y}\,\ud y^2+\epsilon_{z}\,\ud z^2\big)\,.
\end{equation}
We switch from Cartesian to bipolar coordinates using equation \eqref{eq:bipolars}
\begin{eqnarray}
\ud s^2 & = & n_{\scriptscriptstyle \text{IS}}^2 \Bigg\{
    \Bigg[\epsilon_{x}\bigg(\dfrac{\ud x}{\ud \sigma'}\bigg)^2+\epsilon_{y}\,\bigg(\dfrac{\ud y}{\ud \sigma'}\bigg)^2+\epsilon_{z}\,\bigg(\dfrac{\ud z}{\ud \sigma'}\bigg)^2\Bigg]\,\ud \sigma'^2 \nonumber \\
& & \quad\:\,+ \Bigg[\epsilon_{x}\bigg(\dfrac{\ud x}{\ud \tau}\bigg)^2+\epsilon_{y}\,\bigg(\dfrac{\ud y}{\ud \tau}\bigg)^2+\epsilon_{z}\,\bigg(\dfrac{\ud z}{\ud \tau}\bigg)^2\Bigg]\,\ud \tau^2 + \epsilon_{z}\,\ud z^2 \Bigg \}.
\end{eqnarray}
Again, we define the expressions in square brackets in front of $\ud \sigma'^2$ and $\ud \tau^2$ as $\epsilon_{\sigma'}$ and $\epsilon_{\tau}$ respectively and obtain
\begin{equation}
\ud s^2=n_{\scriptscriptstyle \text{IS}}^2\big(\epsilon_{\sigma'}\,\ud \sigma'^2+\epsilon_{\tau}\,\ud \tau^2+\epsilon_{z}\,\ud z^2\big)\,.
\end{equation}
Finally, we change from $\sigma'$ to $\sigma$
\begin{equation}
\ud s^2=n_{\scriptscriptstyle \text{IS}}^2\bigg[\epsilon_{\sigma'}\bigg(\dfrac{\ud\sigma'}{\ud\sigma}\bigg)^2\ud \sigma^2+\epsilon_{\tau}\,\ud \tau^2+\epsilon_{z}\,\ud z^2\bigg]\,.
\end{equation}
We define $\epsilon_{\sigma}=\epsilon_{\sigma'}\bigg(\dfrac{\ud\sigma'}{\ud\sigma}\bigg)^2$ and deduce the metric tensor from the above line element
\begin{equation}
g_{ij}=n_{\scriptscriptstyle \text{IS}}^2\,\text{diag}(\epsilon_{\sigma},\epsilon_{\tau},\epsilon_{z})\,.
\end{equation}
The determinant of this tensor is $g=n_{\scriptscriptstyle \text{IS}}^6\epsilon_{\sigma}\epsilon_{\tau}\epsilon_{z}$ and its inverse is given by
\begin{equation}
g^{ij}=\dfrac{1}{n_{\scriptscriptstyle \text{IS}}^2}\,\text{diag}\bigg(\dfrac{1}{\epsilon_{\sigma}},\dfrac{1}{\epsilon_{\tau}},\dfrac{1}{\epsilon_{z}}\bigg)\,.
\end{equation}
The metric tensor of the background geometry is given by
\begin{equation}
\gamma_{jk}=\text{diag}(\gamma_{\sigma},\gamma_{\tau},\gamma_{z})\,.
\end{equation}
The determinant of $\gamma_{jk}$ is $\gamma=\gamma_{\sigma}\gamma_{\tau}\gamma_{z}$, where
\begin{equation} \label{eq:PhysicalBipolars}
\gamma_{\sigma}=\gamma_{\tau}=\dfrac{a^2}{(\text{cosh}\,\tau-\text{cos}\,\sigma)^2}\qquad \text{and} \qquad \gamma_{z}=1\,.
\end{equation}
In the expression above $a$ is the radius of the stretched region of the cloak [see equation \eqref{eq:bipolars}].\\
We can now substitute the relevant expressions into equation \eqref{eq:Recipe} and obtain the permittivity and permeability tensors
\begin{equation}
\varepsilon^{i}_{\:k_{\text{\:outer}}}=\mu^{i}_{\:k}=\text{diag}(\varepsilon_{\sigma},\varepsilon_{\tau},\varepsilon_{z})=n_{\scriptscriptstyle \text{IS}}\,\text{diag}\bigg(\sqrt{\dfrac{\gamma_{\sigma}\epsilon_{\tau}\epsilon_{z}}{\epsilon_{\sigma}\gamma_{\tau}\gamma_{z}}},   \sqrt{\dfrac{\epsilon_{\sigma}\gamma_{\tau}\epsilon_{z}}{\gamma_{\sigma}\epsilon_{\tau}\gamma_{z}}},   \sqrt{\dfrac{\epsilon_{\sigma}\epsilon_{\tau}\gamma_{z}}{\gamma_{\sigma}\gamma_{\tau}\epsilon_{z}}}\,\bigg)\,.
\end{equation}
The permittivity and permeability tensors given above describe the electromagnetic properties of the outer branch of the device. Note that we expressed the material properties in terms of the mixed (co- and contravariant) tensors, since they are directly related to the refractive indices [see equation \eqref{eq:refindex}] and thus have physical significance \cite{Tyc:2008}.\\

\subsubsection{Inner branch} \label{Ap:InnerBranch}

Next, we calculate the material properties of the inner branch of the cloak (red region in Fig.~3A), where $3\pi/4<|\sigma|\le\pi$. This physical region corresponds to the lower sheet of the virtual geometry (Fig.~3B). The lower sheet carries the refractive index profile of Maxwell's Fish Eye $n_{\scriptscriptstyle \text{M}}$ given by 
\begin{equation} \label{eq:FishEye}
n_{\scriptscriptstyle \text{M}}(x',y')=\dfrac{5\cdot2}{1+((x'+a)/(2a))^{2}+(y'/(2a))^2}\text{.}
\end{equation}
Therefore, the line element in bipolar coordinates is given by
\begin{equation}
\ud s^2=n_{\scriptscriptstyle \text{M}}^2\bigg\{\bigg(\dfrac{a^2}{(\text{cosh}\tau-\text{cos}\bar{\sigma})^2}\bigg)(\ud \bar{\sigma}^2+\ud \tau^2)+\ud z^2\bigg\}\,.
\end{equation}
To simplify the line element we define $\lambda_{\bar{ \sigma}}=\lambda_{\tau}=(a^2/(\text{cosh}\tau-\text{cos}\bar{\sigma})^2)$ and obtain
\begin{equation}
\ud s^2=n_{\scriptscriptstyle \text{M}}^2\bigg\{\lambda_{\bar{\sigma}}\ud \bar{\sigma}^2+\lambda_{\tau}\,\ud \tau^2+\ud z^2\bigg\}\,.
\end{equation}
Next we bring the values of the sigma coordinate from the ranges $[-\pi,0]$ and $[0,\pi]$ to $[\pi,2\pi]$ and $[-2\pi,-\pi]$ respectively using the function
\begin{equation}
\sigma'(\bar{\sigma})= \left\{
\begin{array}{lr}
\bar{\sigma}-2\,\pi\qquad & \text{for } \bar{\sigma} > 0\;\:\\
\bar{\sigma}+2\,\pi & \text{for } \bar{\sigma} \le 0\,.
\end{array} \right.
\end{equation}
We obtain
\begin{equation}
\ud s^2=n_{\scriptscriptstyle \text{M}}^2\bigg\{\lambda_{\bar{\sigma}}\bigg(\dfrac{\ud\bar{\sigma}}{\ud\sigma'}\bigg)^2\ud \sigma'^2+\lambda_{\tau}\,\ud \tau^2+\ud z^2\bigg\}\,=n_{\scriptscriptstyle \text{M}}^2\bigg\{\lambda_{\bar{\sigma}}\ud \sigma'^2+\lambda_{\tau}\,\ud \tau^2+\ud z^2\bigg\}\,,
\end{equation}
where the last equality follows from the fact that $(\ud\bar{\sigma}/\ud\sigma')^2=1$.
Finally, we switch back from $\sigma'$ to $\sigma$ to reverse the expansion of space using equation \eqref{eq:SigmaPrime} and obtain
\begin{equation}
\ud s^2=n_{\scriptscriptstyle \text{M}}^2\bigg\{\lambda_{\bar{\sigma}}\bigg(\dfrac{\ud\sigma'}{\ud\sigma}\bigg)^2\ud \sigma^2+\lambda_{\tau}\,\ud \tau^2+\ud z^2\bigg\}\,=n_{\scriptscriptstyle \text{M}}^2\bigg\{\lambda_{\sigma}\ud \sigma^2+\lambda_{\tau}\,\ud \tau^2+\lambda_{z}\ud z^2\bigg\}\,.
\end{equation}
In the expression above the last equality follows from defining the factor in front of $\ud\sigma^2$ as $\lambda_{\sigma}$ and defining $\lambda_{z}=1$.
The metric tensor of the background geometry is given by
\begin{equation}
\kappa_{jk}=\text{diag}(\kappa_{\sigma},\kappa_{\tau},\kappa_{z})\,.
\end{equation}
The determinant of $\kappa_{jk}$ is $\kappa=\kappa_{\sigma}\kappa_{\tau}\kappa_{z}$, where
\begin{equation} \label{eq:PhysicalBipolars2}
\kappa_{\sigma}=\kappa_{\tau}=\dfrac{a^2}{(\text{cosh}\,\tau-\text{cos}\,\sigma)^2}\qquad \text{and} \qquad \kappa_{z}=1\,.
\end{equation}
We substitute the relevant expressions into equation \eqref{eq:Recipe} and thus obtain
\begin{equation}
\varepsilon^{i}_{\:k_{\text{\;inner}}}=\mu^{i}_{\:k}=\text{diag}(\varepsilon_{\sigma},\varepsilon_{\tau},\varepsilon_{z})=n_{\scriptscriptstyle \text{IS}}\,\text{diag}\bigg(\sqrt{\dfrac{\kappa_{\sigma}\lambda_{\tau}\lambda_{z}}{\lambda_{\sigma}\kappa_{\tau}\kappa_{z}}},   \sqrt{\dfrac{\lambda_{\sigma}\kappa_{\tau}\lambda_{z}}{\kappa_{\sigma}\lambda_{\tau}\kappa_{z}}},   \sqrt{\dfrac{\lambda_{\sigma}\lambda_{\tau}\kappa_{z}}{\kappa_{\sigma}\kappa_{\tau}\lambda_{z}}}\,\bigg)\,.
\end{equation}
Fig.~6 shows the distribution of the tensor eigenvalues within both the inner and outer branches of the combined cloak. For the parameter ranges we get $\varepsilon_{\sigma} \in [1,9.428] $, $\varepsilon_{\tau} \in [1,73.21]$, and $\varepsilon_{z} \in [1,2022.12]$. 

\subsection{Optimization} \label{Ap:Optimization}

In this section we determine the optimal parameters for the topology that combines the partially transmuted Invisible Sphere and Maxwell's Fish Eye Lens (Fig.~3B). We emphasize that this geometry is merely an example, which demonstrates that full invisibility cloaking is possible without superluminal velocities. This geometry could be modified in various ways (e.g. by changing the topology slightly or by introducing further coordinate transformations) to find a set-up with less demanding material parameters.\\
While positioning the branchcut (yellow line in Fig.~9) within the Invisible Sphere (purple circle in Fig.~9), we have to ensure that all index values are raised above one within the stretched region of the cloak and that total internal reflection is avoided within the cloak. Without the background of the Invisible Sphere, the lowest eigenvalue of the stretched region of the cloak (blue region in Fig.~3A) is $0.333$ in the $\varepsilon_{\sigma}$ profile. This value lines the entire length of the branchcut. Therefore, the branchcut has to be placed in the region where $n_{\scriptscriptstyle \text{IS}}>3$ (within the green circle in Fig.~9).\\
\begin{figure}[h!]
\begin{center}
\includegraphics[width=8.5cm]{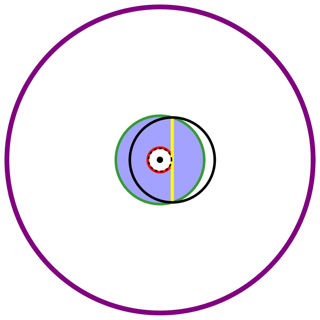}
\caption{The position and size of the branchcut in the Invisible Sphere. The blue annulus highlights the region where the branchcut can be placed such that all tensor eigenvalues are greater than unity and total internal reflection is avoided. The yellow line indicates the orientation and maximal size of the branchcut. The black dashed circle shows the boundary of the transmutation of the Invisible Sphere.}
\end{center}
\end{figure}Total internal reflection may occur when a light ray passes from an optically dense medium to a lighter one. To avoid total internal reflection, we have to ensure that the refractive index along the branchcut is higher in the inner cloak than in the outer cloak. Since the expansion of space was achieved by a smooth function, the expansion by itself would give rise to a smooth index profile. Therefore, total internal reflection will depend solely on the values that the Invisible Sphere and Maxwell's Fish Eye assume on the two sides of the branchcut. The Fish Eye profile changes between 5 and 10 and reaches a value of about 8 at the midpoint of the branchcut. Calculations show that total internal reflection is avoided and the size of the branchcut is maximised when the branchcut is oriented as shown and is placed within the blue annulus in Fig.~9, where the red circle seals off the region where $n_{\scriptscriptstyle \text{IS}}>7.89$. Assuming that the Invisible Sphere has unit radius, we obtain $r_{1}=0.08$ and $r_{2}=0.288$ for the radii of the red and green circles respectively. This yields $0.554$ for the length of the branchcut (i.e. $a=0.277$ in equations (\ref{eq:bipolars}, \ref{eq:complexbipolars}, \ref{eq:inversecomplexbipolars}, \ref{eq:PhysicalBipolars} and \ref{eq:PhysicalBipolars2}).\\
Part of the stretched region (black circle in Fig.~9) will now lie outside the blue annulus. However, calculations show that $\varepsilon_{\sigma}$ goes sufficiently quickly to unity towards the boundary of the stretched region to ensure that the combined index values do not dip below one.\\
Finally, we have to choose a sufficiently small radius for the transmutation of the Invisible Sphere to raise the combined eigenvalues within the transmuted region above one. We found that $b=0.075$ is the greatest and thus optimal transmutation radius that achieves this.\\
Note that refraction will occur at the branchcut when light crosses from the outer to the inner branch due to the differing values of the Invisible Sphere and the Fish Eye profile on the two sides of the branchcut. However, due to the particular symmetry of the device, when the light ray crosses the branchcut for the second time, the effect of the refraction is exactly reversed and thus the trajectory is not distorted overall.\\
We note that, for the purposes of optimization, the refractive index of the inner cloak of our device can be replaced by an index profile that is identical to that of the inner cloak of the Non-Euclidean cloak described in \cite{Leonhardt:2009} and \cite{Tyc:2010} except for an overall multiplicative constant of 5.274. This constant has to be introduced to ensure that all index values in the inner cloak are above unity. Following optimization, we find that the eigenvalues for this construction are $\varepsilon_{\sigma} \in [1,11.65] $, $\varepsilon_{\tau} \in [1,59.2]$, and $\varepsilon_{z} \in [1,944.9]$, which shows that the range of the parameters is already a factor two smaller than the range for the cloak with the Fish Eye profile. However, the size of the invisible region also becomes smaller for this cloak. It is generally true that the greater the invisible region, the higher tensor eigenvalues are needed.


\end{document}